\documentclass[a4paper,final]{appolb}
\usepackage{epsfig}
\usepackage{graphicx} 
\usepackage{amsmath}
\usepackage{amssymb} 
\usepackage{bbm} 
\usepackage[small]{caption2} 
\usepackage{fleqn} 
\usepackage[small,loose]{subfigure}  
\usepackage{mciteW} 

\newcommand{\CenterObject}[1]{\ensuremath{\vcenter{\hbox{#1}}}}
 
\newcommand{\E}[1]{\ensuremath{\mathrm{E}_{#1}}} 
\newcommand{\G}[1]{\ensuremath{\mathrm{G}_{#1}}}
\newcommand{\SO}[1]{\ensuremath{\mathrm{SO}(#1)}}
\newcommand{\SU}[1]{\ensuremath{\mathrm{SU}(#1)}}
\newcommand{\U}[1]{\ensuremath{\mathrm{U}(#1)}}
\newcommand{\Z}[1]{\ensuremath{\mathbbm{Z}_{#1}}} 

\begin{document}
\preprint{DESY 05-260}
\title{
\vspace*{-3cm}
\begin{flushright}
 {\normalfont DESY 05-260}
\end{flushright}
\vspace*{1cm}
\appHuge{Local Grand Unification}%
\thanks{Based on talks presented at the GUSTAVOFEST, Lisbon, July 2005, and 
at the workshop `Strings and the real world', Ohio, November 2005.}%
}
\author{Wilfried Buchm\"uller and Koichi Hamaguchi
\address{Deutsches Elektronen-Synchrotron DESY, 22603 Hamburg, Germany}
\and
Oleg Lebedev and Michael Ratz
\address{Physikalisches Institut der Universit\"at Bonn, Nussallee 12, 53115 Bonn, Germany}
}
\maketitle
\begin{abstract}
\noindent
In the standard model matter fields form complete representations of a grand
unified group whereas Higgs fields belong to incomplete `split' multiplets. This
remarkable fact is naturally explained by `local grand unification' in
higher-dimensional extensions of the standard model. Here, the generations of
matter fields are localized in regions of compact space which are endowed with a
GUT gauge symmetry whereas the Higgs doublets are bulk fields. We realize local
grand unification in the framework of orbifold compactifications of the
heterotic string, and we present an example with \SO{10} as a local GUT group,
which leads to the supersymmetric standard  model as an effective four-dimensional
theory. We also discuss different orbifold GUT  limits and the unification of
gauge and Yukawa couplings. 
\end{abstract}
\preprint{DESY 05-260}
  
\section{Grand unification in $\boldsymbol{D>4}$ and doublet--triplet splitting}

The standard model (SM) is a remarkably successful theory of the structure
of matter. It is based on the gauge group
$G_\mathrm{SM}=\SU3_C\times\SU2_\mathrm{L}\times\U1_Y$ 
and has three generations of matter transforming as
\begin{equation}
        (\boldsymbol{3},\boldsymbol{2})_{1/6}+
        (\overline{\boldsymbol{3}},\boldsymbol{1})_{-2/3}+
        (\overline{\boldsymbol{3}},\boldsymbol{1})_{1/3}+
        (\boldsymbol{1},\boldsymbol{2})_{-1/2}+
        (\boldsymbol{1},\boldsymbol{1})_{1}\;.
\end{equation}
The evidence for neutrino masses strongly supports the existence of right-handed
neutrinos which are singlets under the SM gauge group.  A crucial ingredient of
the SM is further an \SU2 doublet of Higgs fields containing the Higgs boson
which still remains to be discovered. From a theoretical perspective, one would
like to amend the standard model by supersymmetry. Apart from stabilizing the
hierarchy between the electroweak and Planck scales and providing a convincing
explanation of the observed dark  matter, the minimal supersymmetric extension
of the standard model, the MSSM, predicts unification of the gauge couplings at
the unification  scale $M_\mathrm{GUT}\simeq 2\cdot10^{16}\,\mathrm{GeV}$.

Even more than the unification of gauge couplings, the symmetries and the
particle content of the standard model point towards grand unified theories
(GUTs) \cite{Georgi:1974sy,*Pati:1974yy}. Remarkably, one generation of matter,
including  the right-handed neutrino, forms a single spinor representation of
\SO{10} \cite{Georgi:1975qb,*Fritzsch:1974nn}.  It therefore appears natural to
assume an underlying \SO{10} structure of the theory. The route of unification,
continuing via exceptional groups, terminates at $\E8$,
\begin{equation}
 \SO{10} \subset \E6 \subset \E8\;,
\end{equation}
which is beautifully realized in the heterotic string 
\cite{Gross:1984dd,*Gross:1985fr}.

An obstacle on the path towards unification are the Higgs fields, which are 
\SU2 doublets, while the smallest \SO{10} representation containing  Higgs
doublets, the $\boldsymbol{10}$--plet, predicts additional $\SU3_C$ triplets.
The fact that Higgs fields form incomplete `split' GUT  representations is
particularly puzzling in supersymmetric theories where both matter and Higgs
fields are chiral multiplets. The triplets cannot have masses below 
$M_\mathrm{GUT}$ since otherwise proton decay would be too rapid. This then
raises the question why \SU2 doublets are so much lighter than $\SU3_C$
triplets. This is the notorious doublet-triplet splitting problem of ordinary 4D
GUTs.

\begin{figure}[t]
\centerline{\CenterObject{\includegraphics{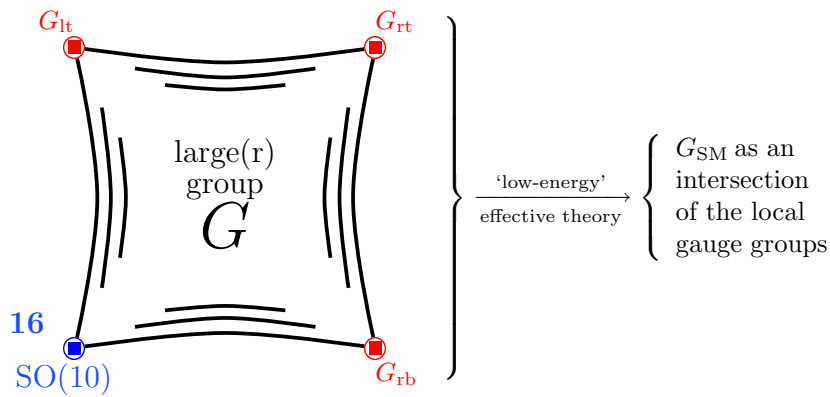}}}
\caption{The picture of local grand unification. The gauge group $G$ is broken
locally to different subgroups. Each of the local subgroups contains the
standard model gauge group $G_\mathrm{SM}$ which emerges as an intersection of
the local groups. `Brane' fields which are confined to a region with certain
symmetry have to come in complete matter multiplets of that symmetry. Hence,
localized $\boldsymbol{16}$-plets of \SO{10} are an attractive explanation of
complete matter generations. Higgs doublets, on the other hand, are states which are
not confined to an \SO{10} region, and can therefore appear as `split
multiplets' in the low--energy spectrum.}
\label{fig:LocalGUTs}
\end{figure}

Higher-dimensional theories offer new possibilities for gauge symmetry breaking
connected with the compactification to four dimensions. A simple and elegant
scheme, leading to chiral fermions in four dimensions, is the compactification
on orbifolds, first considered in string theories 
\cite{Dixon:1985jw,*Dixon:1986jc,Ibanez:1986tp,*Ibanez:1987xa,*Ibanez:1987sn},
and more recently applied to GUT field theories \cite{Kawamura:1999nj,
*Kawamura:2000ev,*Altarelli:2001qj,*Hall:2001pg,*Hebecker:2001wq,*Asaka:2001eh,*Hall:2001xr}. 
In orbifold  compactifications the gauge symmetry of the 4D effective theory is
an intersection of larger symmetries at orbifold fixed points
(cf.~Fig.~\ref{fig:LocalGUTs}). Zero modes located at these fixed points all
appear in the 4D theory and form therefore representations of the larger local
symmetry groups. Zero modes of bulk fields, on the  contrary, are only
representations of the smaller 4D gauge symmetry and form in general `split
multiplets'. Choosing now on some orbifold fixed points \SO{10} as local
symmetry, we obtain the picture of `local grand unification' illustrated in
Fig.~\ref{fig:LocalGUTs}. The SM gauge group is obtained as an intersection of
different local GUT groups. Matter fields appear as $\boldsymbol{16}$--plets
localized at the fixed points, whereas the Higgs doublets are associated with
bulk fields, which provides a solution of the doublet-triplet splitting problem.
In this way the structure of the standard model is naturally reproduced.  

\section{Orbifold compactification of the heterotic string}

\begin{figure}[t]
\centerline{\CenterObject{\includegraphics{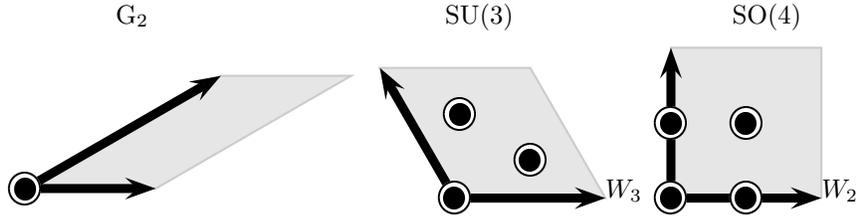}}}
\caption{Root lattice $\Lambda_{G_2\times\SU3\times\SO4}$ and fixed points of
the \Z6 action.}
\label{fig:RootLattice}
\end{figure}

`Local grand unification' naturally occurs in compactifications  of the
$\E8\times\E8$ heterotic string \cite{Gross:1984dd,*Gross:1985fr}. Six of the
ten space-time dimensions are  compactified on an orbifold
\cite{Dixon:1985jw,*Dixon:1986jc,Ibanez:1986tp,*Ibanez:1987sn,Katsuki:1990bf}.
Specifically, we consider $\Z3\times\Z2$ orbifold compactifications on the
lattice $\Lambda_{\G2\times\SU3\times\SO4}$, which is a product of three
two-tori defined by the root vectors of \G2, \SU3 and \SO4, respectively
\cite{Kobayashi:2004ud,*Kobayashi:2004ya}.  The \Z6 action is a rotation by
$2\pi/6$ in the \G2 plane, by $2\pi/3$ in the \SU3 plane and by $2\pi/2$ in the
\SO4 plane. This \Z6 action has fixed points in the various planes as
illustrated in Fig.~\ref{fig:RootLattice}.

The rotation in the compact dimensions is embedded into the gauge degrees of
freedom. It acts as a shift, denoted by $V_6$, on the left-moving coordinates
$X_\mathrm{L}^I$ ($1\le I \le 16$), i.e.\ upon a rotation in the internal
space-time $X_\mathrm{L}^I$ transforms as
$X_\mathrm{L}^I\to X_\mathrm{L}^I+V_6^I$. This shift satisfies
$6V_6\in\Lambda_{\E8\times\E8}$, where $\Lambda_{\E8\times\E8}$ denotes the root
lattice of $\E8\times\E8$, reflecting that the corresponding automorphism has
order 6. In addition, some tori carry Wilson lines. For instance, a torus
translation along one axis in the \SO4 torus is associated with a shift by $W_2$
which has degree two, i.e.\  $2W_2\in\Lambda_{\E8\times\E8}$.  A more detailed
discussion can be found elsewhere
\cite{Forste:2004ie,Kobayashi:2004ud,*Kobayashi:2004ya,Buchmuller:2004hv,Buchmuller:2005pr}.

\subsection{Orbifold construction kit}
\label{sec:ock}

The orbifold action leads to local gauge symmetry breakdown at the fixed points.
To see this, one analyzes locally the invariance conditions for the gauge fields
$A_\mu^p$, corresponding to the generator $p\in\Lambda_{\E8\times\E8}$.  For
instance, at the origin in Fig.~\ref{fig:RootLattice}, the invariance condition
requires that $A_\mu^p$ vanish unless the corresponding generator is
`orthogonal' to the local shift $V_\mathrm{local}=V_6$. This implies that any
gauge boson not fulfilling $p\cdot V_6\equiv0$\footnote{Here `$\equiv$' means
`up to integers'.} is projected out of the zero-mode spectrum. The remaining
gauge bosons form a sub-algebra of the original $\E8\times\E8$. One can
thus say that the origin carries a local gauge group.

Repeating the analysis at other fixed points leads in general to different local
projection conditions. For instance, the local projection condition at the
origin in the \G2 and \SU3 tori and at the bottom right position in the \SO4
torus is the same as before except that the shift now gets amended by the Wilson
line $W_2$, i.e.\ $V_\mathrm{local}=V_6+W_2$. This modified projection
condition, $p\cdot V_\mathrm{local}\equiv0$, implies a different local gauge
group. 
An analogous analysis can be carried out for the remaining fixed points. The
result is that each of the fixed points carries a local gauge group, and those
local gauge groups are in general different. If two local shifts are equal,
i.e.\ if the Wilson line corresponding to the translation connecting the two
fixed points vanishes, the local gauge groups and their embedding into
$\E8\times\E8$ coincide. 

The gauge boson zero-modes consist of the gauge bosons surviving all
local projection conditions simultaneously. In other words, the gauge group
after compactification is an intersection of the local gauge groups. In the
following, we shall focus on the possibility that this intersection is, up to
\U1 factors and a `hidden sector gauge group', the SM gauge group
$G_\mathrm{SM}$ . It is crucial to note that each of the local gauge groups
contains $G_\mathrm{SM}$ as a  subgroup, i.e.\
\begin{equation}
 G_\mathrm{SM}\,\subsetneq\, G_\mathrm{local}\;.
\end{equation} 
This leads to the picture of `local grand unified theories' where
$G_\mathrm{SM}$ emerges as an intersection of different GUT groups residing at
different fixed points.

\begin{figure}[t]
\centerline{
\subfigure[Orbifold construction kit.\label{fig:OrbifoldConstructionKit}]{%
        \CenterObject{\includegraphics{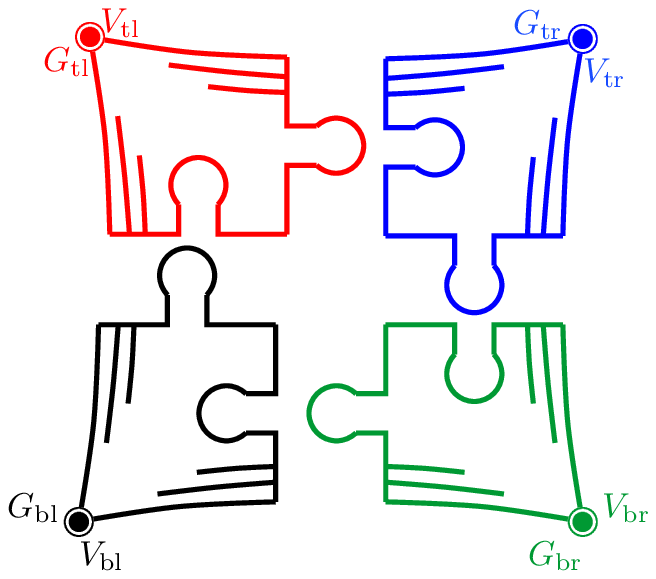}}}      
\quad
\subfigure[$2+1$ family models.\label{fig:2+1}]{
\CenterObject{\includegraphics{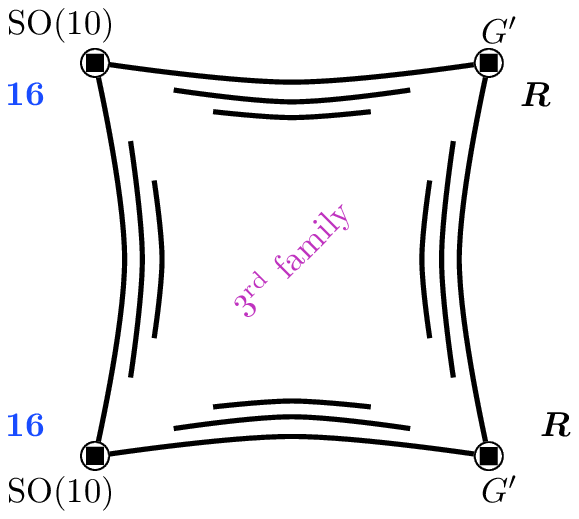}}}
}
\caption{(a) orbifolds can be constructed by combining `corners' carrying a
local gauge group emerging from the action of a local shift. (b) In $2+1$ family
models, two families appear as $\boldsymbol{16}$--plets residing on fixed points
with local \SO{10} symmetry. The third family comes from elsewhere.
}
\end{figure}

The geometry of a 2D orbifold can be visualized as follows
(cf.~\cite{Quevedo:1996sv,*Hebecker:2003jt}). The fundamental region of a
$\Z{N}$ orbifold is one $N^\mathrm{th}$ of the fundamental region of the torus.
One can fold it and identify the remaining edges.  This yields a `pillow' or
`ravioli' where the orbifold fixed points correspond to the corners (cf.\
Fig.~\ref{fig:LocalGUTs}). 

The corners of the pillow serve as building blocks for the construction of an
orbifold model. One starts  at one corner with a local shift leading to a local
gauge group. The simplest possibility is to combine identical corners  which
leads to an orbifold model without Wilson lines.
In order to construct realistic orbifold models, one has to consider
non-vanishing Wilson lines, i.e.\ combine corners with different gauge groups.  
Gluing these corners together leads to an orbifold model with Wilson lines (cf.\
Fig.~\ref{fig:OrbifoldConstructionKit}). 
Let us  emphasize that one cannot combine these corners arbitrarily. Rather,
there are severe constraints coming from modular invariance, which  restrict the
allowed  Wilson lines  (cf.~\cite{Forste:2004ie}).

\subsection{Localized $\boldsymbol{16}$--plets and 2+1 family models}
\label{sec:2+1}

The `orbifold construction kit' described above is a helpful tool for model
building. Note that each local GUT leads to local GUT representations. Among
the zero modes, the representations of the first twisted sector  $T_1$ play 
a special role. They correspond to `brane' fields which are completely 
localized at the fixed points. In particular, they only feel the local gauge 
group and therefore are forced to furnish complete GUT representations. 

An application of this observation to localized $\boldsymbol{16}$--plets of
local \SO{10} GUTs leads to a simple recipe for the construction of three
generation models. One starts with a `corner' which carries a local \SO{10} and
a $\boldsymbol{16}$--plet in the $T_1$ sector. This $\boldsymbol{16}$--plet will
appear as a complete multiplet in the massless spectrum, even though the
low-energy gauge group will be an intersection of \SO{10} with other groups, and
therefore only some subgroup of \SO{10}. As discussed above, this provides
understanding of the fact that there are complete generations and split
multiplets at the same time.

In the geometry introduced above there are only two shifts which produce a
local \SO{10} together with the $\boldsymbol{16}$--plets 
\cite{Katsuki:1989qz,*Katsuki:1989cs},
\begin{eqnarray}
 V_6
 & = & 
 \left(\frac{1}{2},\frac{1}{2},\frac{1}{3},0,0,0,0,0\right)
 \left(\frac{1}{3},0,0,0,0,0,0,0\right)\;,
 \nonumber\\
 V_6'
 & = & 
 \left(\frac{1}{3},\frac{1}{3},\frac{1}{3},0,0,0,0,0\right)
 \left(\frac{1}{6},\frac{1}{6},0,0,0,0,0,0\right)\;.
\end{eqnarray}

In the following, we shall focus on $2+1$ family models, where  two families
come from two \SO{10} corners while the third one comes from somewhere else
(cf.\ Fig.~\ref{fig:2+1}). This pattern has recently been explored in the
context of Pati-Salam models \cite{Kobayashi:2004ud,*Kobayashi:2004ya}. In
contrast to models with three families of localized $\boldsymbol{16}$--plets, in
$2+1$ family models the third family and the Higgs originate partially from the
untwisted sector. As we shall see, this leads naturally to a situation where the
top Yukawa coupling is related to the gauge coupling. Note that an untwisted
$\boldsymbol{16}$--plet, i.e. a bulk third family, occurs only in the
$\E8\times\E8$ heterotic string and not in the \SO{32} heterotic string.


\begin{figure}[t]
\centerline{\CenterObject{\includegraphics{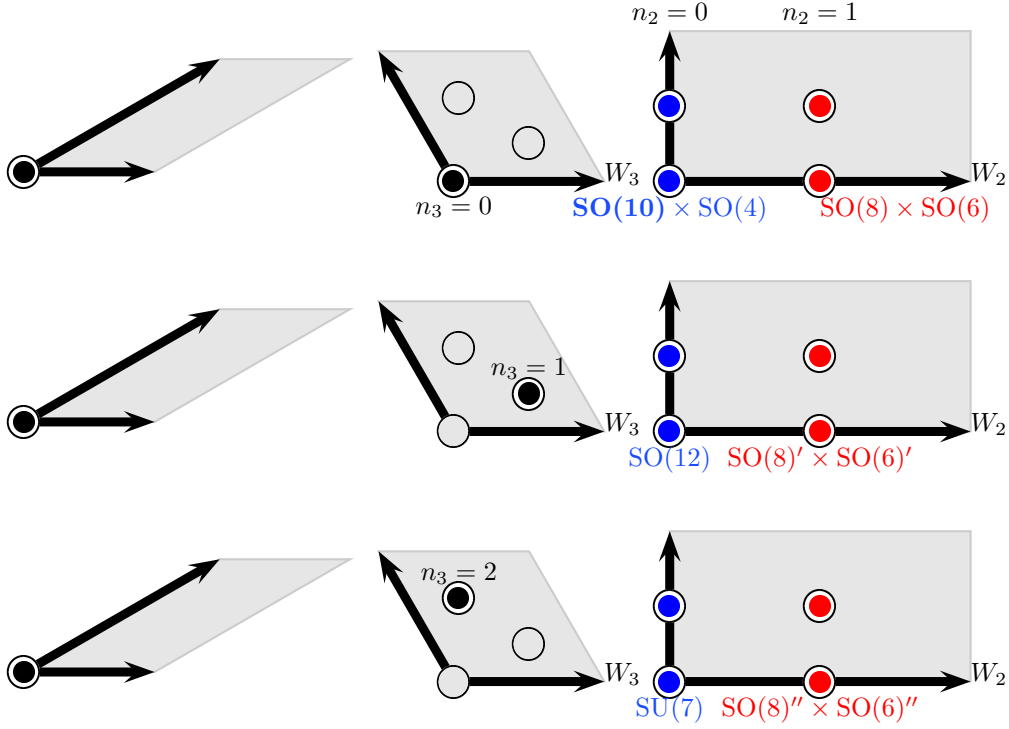}}}
\caption{Local gauge groups up to \U1 factors and subgroups of the second \E8.
As indicated, the fixed point come in six pairs.}
\label{fig:LocalGaugeGroups}
\end{figure}


\section{The MSSM from the heterotic string}
\label{sec:MSSM}

\subsection{Model definition and gauge groups}

Let us now discuss a specific example \cite{Buchmuller:2005jr}. We consider a
compactification of the $\E8\times\E8$ heterotic string on the $\Z3\times\Z2$
orbifold described above. In order to obtain a $2+1$ family model, we
require only two Wilson lines $W_2$ and $W_3$, while a possible third Wilson 
line associated with the  `vertical' translation in the \SO4 torus 
(cf.~Fig.~\ref{fig:RootLattice}) is set to zero. In an orthonormal basis,
the shift and the Wilson lines are given by
\begin{eqnarray}
 V_6 & = &
 \left(\frac{1}{2},\frac{1}{2},\frac{1}{3},0,0,0,0,0\right) \, 
 \left(\frac{1}{3},0,0,0,0,0,0,0\right) 
 \;,\nonumber\\
 W_2 & = & 
 \left(\frac{1}{2},0,\frac{1}{2},\frac{1}{2},\frac{1}{2},0,0,0\right) 
 \,\left(-\frac{3}{4},\frac{1}{4},\frac{1}{4},-\frac{1}{4},\frac{1}{4},\frac{1}{4},\frac{1}{4},-\frac{1}{4}\right) 
 \;,\nonumber\\
 W_3 & = &
 \left(\frac{1}{3},0,0,\frac{1}{3},\frac{1}{3},\frac{1}{3},\frac{1}{3},\frac{1}{3}\right) \, 
 \left(1,\frac{1}{3},\frac{1}{3},\frac{1}{3},0,0,0,0\right)
 \;. 
\end{eqnarray}
The twelve fixed points depicted in Fig.~\ref{fig:LocalGaugeGroups} come in six
inequivalent pairs. They carry various local gauge groups whose intersection,
which is the unbroken gauge group after compactification, is given by
\begin{equation}
 G~=~G_\mathrm{SM}\times\left[\SO{6}\times\SU2\right]\times\U1^8\;.
\label{eq:G}
\end{equation}
Here, the brackets indicate a subgroup of the second \E8. One of the $\U1$
factors is anomalous \cite{Dine:1987xk}. As a consequence, some fields charged
under the anomalous \U1 attain vacuum expectation values (VEVs) 
which break this
\U1  (cf.~\cite{Cleaver:1997jb}).  In our model there exist no 
fields which are charged only under the anomalous \U1.\footnote{This feature 
occurs rather generally in orbifold models with Wilson lines.} Hence, 
further \U1 factors are necessarily broken, which is phenomenologically
attractive. 

\subsection{The massless spectrum}

The zero modes of our model are given in Tab.~\ref{tab:Spectrum}, labeled 
by their quantum numbers w.r.t. the gauge group 
$G_\mathrm{SM}\times[\SO{6}\times\SU2]$. Further details such as the oscillator 
numbers will be presented elsewhere \cite{Buchmuller:2005pr}. 
Here we only highlight some aspects of the spectrum.

\begin{table}[t]
\centerline{
 \begin{tabular}{|c|c|c|c|c|c|c|}
 \hline
 name & irrep & count & &
 name & irrep & count\\
 \hline
 $q_i$ & $(\boldsymbol{3},\boldsymbol{2};\boldsymbol{1},\boldsymbol{1})_{1/6}$ & 3 & &
 $\bar u_i$ & $(\overline{\boldsymbol{3}},\boldsymbol{1};\boldsymbol{1},\boldsymbol{1})_{-2/3}$ 
        & 3\\
 $\bar d_i$ & $(\overline{\boldsymbol{3}},\boldsymbol{1};\boldsymbol{1},\boldsymbol{1})_{1/3}$ 
        & 7 & & 
 $d_i$ & $(\boldsymbol{3},\boldsymbol{1};\boldsymbol{1},\boldsymbol{1})_{-1/3}$ & 4\\
 $\bar\ell_i$ & $(\boldsymbol{1},\boldsymbol{2};\boldsymbol{1},\boldsymbol{1})_{1/2}$ & 5 & &
 $\ell_i$ & $(\boldsymbol{1},\boldsymbol{2};\boldsymbol{1},\boldsymbol{1})_{-1/2}$ & 8\\
 $m_i$ & $(\boldsymbol{1},\boldsymbol{2};\boldsymbol{1},\boldsymbol{1})_{0}$ & 8 & &
 $\bar e_i$ & $(\boldsymbol{1},\boldsymbol{1};\boldsymbol{1},\boldsymbol{1})_{1}$ & 3 \\
 $s^-_i$ & $(\boldsymbol{1},\boldsymbol{1};\boldsymbol{1},\boldsymbol{1})_{-1/2}$ & 16 & & 
 $s^+_i$ & $(\boldsymbol{1},\boldsymbol{1};\boldsymbol{1},\boldsymbol{1})_{1/2}$ & 16\\
 $s^0_i$ & $(\boldsymbol{1},\boldsymbol{1};\boldsymbol{1},\boldsymbol{1})_{0}$ &
 69
 & & $h_i$ &
 $(\boldsymbol{1},\boldsymbol{1};\boldsymbol{1},\boldsymbol{2})_{0}$& 14\\
 $f_i$ & $(\boldsymbol{1},\boldsymbol{1};\boldsymbol{4},\boldsymbol{1})_{0}$ & 4 & &
 $\bar f_i$ &
 $(\boldsymbol{1},\boldsymbol{1};\overline{\boldsymbol{4}},\boldsymbol{1})_{0}$ & 4 \\
 $w_i$ & $(\boldsymbol{1},\boldsymbol{1};\boldsymbol{6},\boldsymbol{1})_{0}$ &
 5 & & & &\\
 \hline
 \end{tabular}  
}
\caption{The $G_\mathrm{SM}\times[\SO{6}\times\SU2]$ quantum numbers
of the spectrum. The hypercharge is indicated by the subscript.}
\label{tab:Spectrum} 
\end{table} 

A key ingredient is the appearance of the `local GUTs'.  In
Tab.~\ref{tab:LocalGUTs} we list the local gauge groups together with the local
$T_1$ representations. As discussed in Section \ref{sec:2+1} our shift $V_6$ is
chosen so as to  produce two  local $\boldsymbol{16}$--plets at 
$n_2=n_3=0,\ n_2'=0,1$. Note that these are the only $T_1$ fields at $n_2=0$ 
which transform non-trivially under $G_\mathrm{SM}$. At $n_2=1$, there are
$\boldsymbol{4}$--plets and $\overline{\boldsymbol{4}}$--plets w.r.t.\ \SO6
subgroups of the first \E8. Although these \SO6 subgroups are 
embedded into \E8  differently   for different $n_3$, each $\boldsymbol{4}$--plet or
$\overline{\boldsymbol{4}}$--plet gives rise to an $\SU2_\mathrm{L}$ doublet
with zero hypercharge ($m_i$ of Tab.~\ref{tab:Spectrum}) and two non--Abelian
singlets with opposite hypercharge ($s^\pm_i$ of Tab.~\ref{tab:Spectrum}). In
particular, apart from the two  $\boldsymbol{16}$--plets the $T_1$
states are vector--like w.r.t.\ $G_\mathrm{SM}$.

\begin{table}[t]
\centerline{\begin{tabular}{|l|l|l|}
\hline
& $n_2=0$ & $n_2=1$ \\
\hline
$n_3=0$ 
& 
$\SO{10}\times\SU2^2\times\left[\SO{14}\right]$ 
& 
$\SO8\times\SO6\times\left[\SU7\right]$
\\
& 
$\left(\boldsymbol{16},\boldsymbol{1},\boldsymbol{1};\boldsymbol{1}\right)$
& 
$\left(\boldsymbol{1},\boldsymbol{4};\boldsymbol{1}\right)$ 
\\
& 
$\oplus\ 2\times 
\left(\boldsymbol{1},\boldsymbol{2},\boldsymbol{1};\boldsymbol{1}\right)
\oplus\left(\boldsymbol{1},\boldsymbol{1},\boldsymbol{2};\boldsymbol{1}\right)$
& 
\\
\hline
$n_3=1$ & $\SO{12}\times\left[\SO8\times\SO6\right]$ & 
        $\SO8'\times\SO6'\times\left[\SU7\right]$\\
 & $\left(\boldsymbol{1};\boldsymbol{8},\boldsymbol{1}\right)
 \oplus \left(\boldsymbol{1};\boldsymbol{1},\boldsymbol{4}\right)$      
 & $\left(1,\boldsymbol{4};\boldsymbol{1}\right)$ \\
\hline 
$n_3=2$ & $\SU7\times\left[\SO8\times\SO6\right]$ 
        & $\SO8''\times\SO6''\times\left[\SO{10}\times\SU2^2\right]$\\
 & $\left(\boldsymbol{1};\boldsymbol{1},\overline{\boldsymbol{4}}\right)$ 
 & $\left(\boldsymbol{1},\overline{\boldsymbol{4}};\boldsymbol{1},
        \boldsymbol{1},\boldsymbol{2}\right)$   \\
\hline
\end{tabular}}
\caption{Local GUTs and local  representations. The
geometric interpretation of the `localization quantum numbers' $n_2$ and $n_3$
is given in Fig.~\ref{fig:LocalGaugeGroups}.}
\label{tab:LocalGUTs}
\end{table}

To understand the origin of the third generation, recall that the shift $V_6$
breaks the first $\E8$ to $\SO{10}\times\SU2^2\times\U1$. Under this breaking,
some internal components of the gauge bosons transforming in the coset
$\E8/[\SO{10}\times\SU2^2\times\U1]$ remain massless. An important property of
this coset  is that it contains
$\boldsymbol{16}$--plets. In the $U_1$ and  $U_2$ sectors we obtain
\SO{10} $\boldsymbol{16}$--plets, while in the $U_3$ sector we obtain \SO{10}
$\boldsymbol{10}$--plets. Here, the $U_1$, $U_2$ and $U_3$ sectors consist of 
the gauge bosons with spatial
 components in the direction of the \G2, \SU3, and \SO4
torus, respectively. Due to the  presence of Wilson lines, these 
\SO{10} $\boldsymbol{16}$--plets and $\boldsymbol{10}$--plets are
subject to further projections  such that we finally obtain
$(\overline{\boldsymbol{3}},\boldsymbol{1})_{-2/3}$ and
$(\boldsymbol{1},\boldsymbol{1})_{1}$ w.r.t.\ $G_\mathrm{SM}$ from $U_1$, and
$(\boldsymbol{3},\boldsymbol{2})_{1/6}$ from $U_2$. These states correspond  to
a $\boldsymbol{10}$--plet of \SU5.
Together with localized states from   higher
twisted sectors, one  obtains an additional complete generation. This
seemingly miraculous completion of the third generation is related to  the
necessary SM anomaly cancellation. 

In summary, the massless spectrum contains three generations, two of which
originate from the $T_1$ sector and one being a mixture of states from
$U$, $T_2$ and $T_4$. All other states are \emph{vector-like} w.r.t.\ 
$G_\mathrm{SM}$. Thus we have
\begin{equation}  
 \text{matter:}~~~~3\times\boldsymbol{16} ~~+~~ \text{vector-like} \;.
\end{equation}  

One of the main problems of most known string models is the presence of exotic
states at low energies. In our model the vector--like  exotic states
do not appear at low energies. Their mass terms are
generated by the  superpotential involving  singlet fields and consistent with string selection rules
\cite{Hamidi:1986vh,*Dixon:1986qv,Font:1988tp,*Font:1988mm},
\begin{equation} 
 W~\supset~x_i \bar x_j \langle  s_a^0 s_b^0 \dots   \rangle \;.
\end{equation}
The singlet fields  acquire large VEVs yielding  masses for all
exotic states \cite{Buchmuller:2005jr,Buchmuller:2005pr}. Recently, an 
interesting class of heterotic string compactifications has been constructed 
where massless exotic states are absent from the beginning 
\cite{Braun:2005ux,*Bouchard:2005ag,*Braun:2005nv}.
Their  relation to the model described above remains to be clarified.

\begin{figure}[t]
\centerline{\CenterObject{\includegraphics{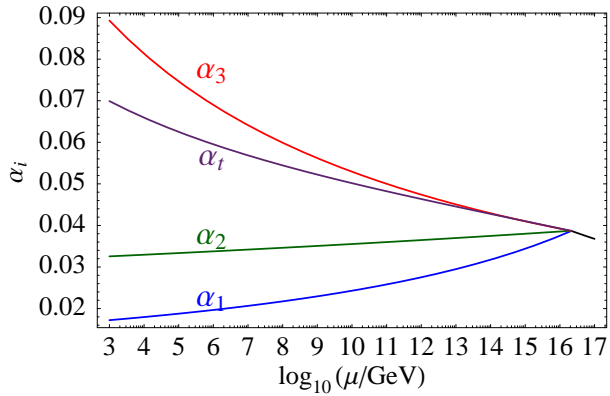}}}
\caption{Illustration of gauge--Yukawa unification. The plot shows the MSSM
evolution of $\alpha_i=g_i^2/(4\pi)$ and $\alpha_t=y_t^2/(4\pi)$ where $g_i$
denotes the gauge couplings and $y_t$ is the top Yukawa coupling.}
\label{fig:GaugeYukawaUnification}
\end{figure}

\subsection{Unification of couplings and flavour structure}

Our model has no exotic states at low energies and admits TeV scale soft
masses. 
Therefore, it is consistent with gauge coupling unification.
Furthermore, the top--Yukawa coupling is related to the gauge
coupling since the third generation originates partially from the untwisted
sector. As described above, the $U_3$ zero modes descend from an 
\SO{10} $\boldsymbol{10}$--plet, and give rise to two states $\ell_1$ and 
$\bar\ell_1$ with the quantum numbers of the MSSM Higgs doublets. The 
up--quark and the quark doublet of the third generation come from the
untwisted sectors $U_1$ and $U_2$, respectively. The Yukawa coupling 
$U_1\,U_2\,U_3$ is a gauge interaction in 10D and its strength
is given by the gauge coupling at the unification scale.
One thus obtains the 
superpotential coupling
\begin{equation}
 W~\supset~y_t\,q_3\,\bar u_3\,h_u\;,
\end{equation}
with $y_t\simeq g$ at the GUT scale, as long as the MSSM `up--type' Higgs is 
dominated by the $U_3$ state, $h_u\simeq\bar\ell_1$ 
(cf.~Fig.~\ref{fig:GaugeYukawaUnification}). It is well known that this is 
consistent with the measured top mass. The  Yukawa couplings involving
the first two generation up--type quarks occur only through higher
dimensional operators \cite{Buchmuller:2005jr}, which leads to the 
 required suppression of these couplings.

The light down--type quarks $\bar d_i$ and lepton doublets $\ell_i$ are  linear
combinations of states from various twisted and untwisted sectors. A similar
setup has recently been proposed in the context  of orbifold GUTs
\cite{Asaka:2003iy}. There, it has been shown that a mixing  of the light
leptons (down--type  quarks) with additional heavy  leptons 
(down--type quarks) can lead to the observed large  neutrino mixings
together with small CKM mixings in the quark sector. In our  heterotic
string model an analogous mixing occurs. 
A more
detailed analysis of  phenomenological aspects of the model will be presented
elsewhere.

\begin{figure}[t]
\centerline{\subfigure[\SO4 plane large.\label{fig:OGL1}]{%
\CenterObject{\includegraphics{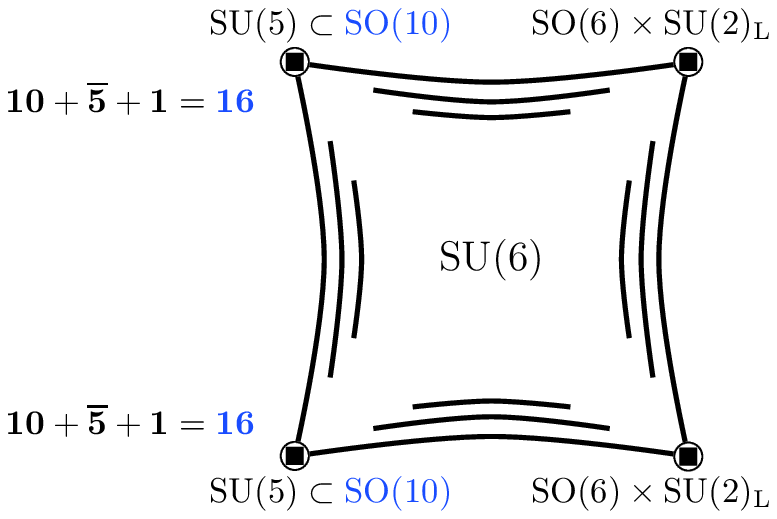}}}
\quad
\subfigure[\SU3 plane large.\label{fig:OGL2}]{%
\CenterObject{\includegraphics{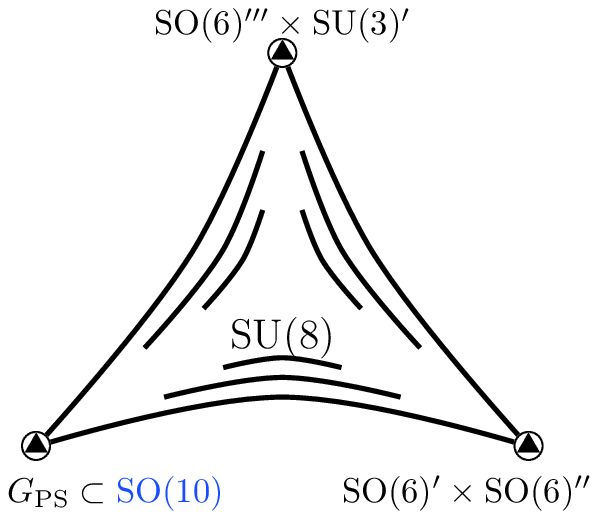}}}}
\centerline{\subfigure[\G2 plane large.\label{fig:OGL3}]{%
\CenterObject{\includegraphics{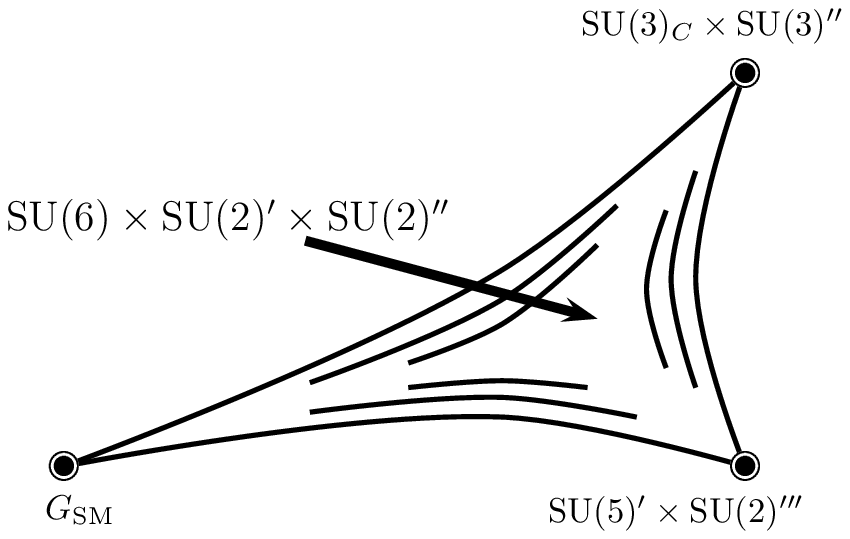}}}}
\caption{Orbifold GUT limits. Only subgroups of the first \E8 are shown, and \U1
factors are omitted.}
\label{fig:OGL}
\end{figure}

\subsection{Orbifold GUT limits}

One of the main motivations for revisiting orbifold compactifications of the
heterotic string is the phenomenological success of orbifold GUTs. It is
therefore interesting to study orbifold GUT limits of the model, which 
correspond to anisotropic compactifications where some radii are significantly
larger than the others. Such  anisotropy may mitigate the discrepancy 
between the GUT and the string scales \cite{Witten:1996mz,*Hebecker:2004ce}.
In the energy range between the compactification scale and the string scale 
one  obtains an effective higher-dimensional field theory.

We focus on $4+2$--dimensional orbifold GUT limits, and keep only subgroups 
of the first \E8 factor in the subsequent analysis. As an example, 
consider the case that the  compactification  radii of the \SO4 plane are larger than the others.
For energies in the range $R_{\SO4}^{-1}<E<R_{\G2,\SU3}^{-1}$ the model is
described by an effective six--dimensional theory. The local gauge groups
at the fixed points and the gauge group in the bulk are obtained by imposing
the following projection conditions on the gauge bosons labeled by $p$:
\begin{align}
 \text{bulk} & :&&  2V_6\cdot p~\equiv~0\;, & W_3\cdot p~\equiv~0\;,
 \nonumber\\
 n_2=0 & :&&  V\cdot p ~\equiv~0\;,& W_3\cdot p~\equiv~0\;,
 \nonumber\\
 n_2=1 & : && (V+W_2)\cdot p ~\equiv~0\;, & W_3\cdot p~\equiv~0\;.
\end{align}
This leads to the bulk group \SU6 and
to the `local GUTs' \SU5 at $n_2=0$ and $\SO6\times\SU2_\mathrm{L}$ at $n_2=1$,
respectively (cf.\ Fig.~\ref{fig:OGL} (a)). The two fixed points with $n_2=0$
carry one generation each. The appearance of these complete generations cannot
be fully understood in the 6D orbifold GUT limit. To have  deeper 
understanding one has to zoom into the smaller dimensions and unravel the 
underlying \SO{10} gauge symmetry. An important property of this limit is that
the bulk \SU6 is a simple group containing $G_\mathrm{SM}$. Hence
the running of the gauge
couplings in six dimensions does not discriminate between the subgroups of
$G_\mathrm{SM}$. This supports the picture of gauge coupling unification at the
compactification scale.

An analogous analysis can be carried out when the \SU3 torus is larger than the
other tori. In this case, one obtains the orbifold GUT shown in Fig.~\ref{fig:OGL} (b) 
as an effective theory. The first two generations are localized at the fixed 
point with the Pati--Salam group 
$G_\mathrm{PS}=\SO6\times\SU2_\mathrm{L}\times\SU2_\mathrm{R}$
as an unbroken local GUT. Again, the bulk
group is simple and contains all SM group factors.

In the limit where the \G2 torus  is large, the first two generations reside at
the fixed point with unbroken gauge symmetry $G_\mathrm{SM}$ 
(cf.~Fig.~\ref{fig:OGL} (c)). In this case, $\SU3_C\times\SU2_\mathrm{L}$ is a
subgroup of the bulk group $\SU6$, but hypercharge is not fully contained in
this \SU6. However, this does not lead to the running or threshold corrections 
which discriminate between the hypercharge and the simple SM subgroups because
of $N=4$ supersymmetry in the bulk, implying vanishing 
$\beta$--functions.

In summary, we find that all orbifold limits are consistent with gauge coupling
unification up to possible logarithmic corrections coming from states localized
at the fixed points. The different geometries have, however, important 
consequences for the values of the gauge couplings at the compactification
scale and also for the pattern of the Yukawa couplings. These issues will be 
analyzed in more detail elsewhere.

\section{Summary}

The quest for unification is a central theme of particle physics. In a
bottom--up approach, starting from the standard model, one is first
led to conventional 4D GUTs. 
Among the GUT groups, \SO{10} is
singled out since one matter generation, including
the right--handed neutrino, forms an  irreducible
representation of  \SO{10}, the $\boldsymbol{16}$--plet.  The route of
unification, continuing via exceptional groups, terminates at \E8.

An attractive starting point for a unified theory including gravity
is, in a top--down approach, the heterotic string with the gauge group
$\E8\times\E8$. 
In its orbifold compactifications, GUT groups appear locally  at orbifold
fixed points.
We have
presented an example with local \SO{10} symmetry and localized
$\boldsymbol{16}$--plets from the twisted sectors. The standard model
gauge group is  obtained as an intersection of different local \E8
subgroups. The model has three quark--lepton generations, one pair
of Higgs doublets and no exotic matter.

\vspace*{0.5cm}\noindent
{\bf Acknowledgements.} It is a pleasure to thank T.~Kobayashi, M.~Lindner,
J.~Louis, H.~P.~Nilles and S.~Stieberger for valuable discussions. 
W.~B. thanks the organizers of the GUSTAVOFEST for arranging an enjoyable
and stimulating Symposium in Honor of Gustavo~C.~Branco. 
M.~R. is deeply indebted to A.~Wisskirchen for technical support. This work was
partially supported by the European Union 6th Framework Program
MRTN-CT-2004-503369 ``Quest for Unification'' and MRTN-CT-2004-005104
``ForcesUniverse''.

\bibliography{Orbifold}
\addcontentsline{toc}{section}{Bibliography}
\bibliographystyle{ArXivmciteW}

\end{document}